\pdfoutput=1  
\documentclass[3p,twocolumn]{elsarticle}  

\usepackage{graphicx}
\usepackage{booktabs}
\usepackage{multirow}
\usepackage{amssymb}
\usepackage{hyperref}  

\biboptions{numbers,sort&compress}

\makeatletter
\def\ps@pprintTitle{%
  \let\@oddhead\@empty
  \let\@evenhead\@empty
  \let\@oddfoot\@empty
  \let\@evenfoot\@oddfoot}
\makeatother

\begin{document}

\begin{frontmatter}

\title{Fast Unsupervised Anomaly Detection Methods on Software Logs - An Empirical Study}

\author[inst1]{Jesse Nyyssölä\corref{cor1}}
\ead{jesse.nyyssola@helsinki.fi}

\author[inst1]{Mika Mäntylä}
\ead{mika.mantyla@helsinki.fi}

\author[inst1]{Yuqing Wang}
\ead{yuqing.wang@helsinki.fi}

\cortext[cor1]{Corresponding author}

\affiliation[inst1]{organization={Department of Computer Science, University of Helsinki},
            city={Helsinki},
            country={Finland}}
\begin{abstract}

\textbf{Context:}
Anomaly detection on software logs ensures system reliability through failure identification, root cause analysis and fault localization. While supervised models show excellent performance, in a practical setting there rarely are labeled training data available. This calls for the use of unsupervised models. 

\textbf{Objectives:}
We study effects of data preprocessing, anomaly detection method, and the dataset. For preprocessing, we examine three options: The use of anomaly-free training data, using randomized as opposed to chronologically selected training data, and the use of a custom loader to extract log message from the log line. Regarding anomaly detection we use four models (KMeans, IsolationForest, RarityModel, Out-Of-Vocabulary Detector) and three representations (Words, trigrams, and parsed events). We evaluate these approaches on seven datasets: BGL, HDFS, Hadoop, Spirit, Thunderbird, Liberty, and LO2Token. 

\textbf{Methods:}
We run each combination of the selected factors and measure performance with AUCROC and F1-score metrics. Then we utilize a generalized linear model to establish relationships between the factors. Furthermore, we address execution times. 

\textbf{Results:}
Our evidence reveals large variability across all datasets, preventing us from making recommendations that apply universally. For instance, log parsing enhances performance on some datasets but reduces it on others.
The evidence suggests that using anomaly-free data for training unsupervised models can improve anomaly detection performance, though this requires extra effort to identify such data.
Previous work has argued that shuffling data leads to unrealistic model performance (data leakage). Yet, our results show this does not hold for the unsupervised models we tested, while it remains true for supervised models in the same setup.
Regarding execution time, the results are clear. For fast execution of the pipeline, use word representation and avoid IsolationForest model. 

\textbf{Conclusion:}
We encourage practitioners to be aware of dataset-specific differences, and study what has worked for similar datasets in the past.


\end{abstract}


\end{frontmatter}

\section{Introduction}

Software log anomaly detection has two contrasting approaches. First is based on deep-learning, which is emphasized on research publications \cite{literaturereview}, \cite{surveydl}, while the second is based on simple rule-based detectors and often used in industry \cite{adhasurvey}.
Computationally intensive deep-learning approaches, while gaining performance improvements, are practically harder to utilize in practice due to their complexity, heavy computational cost, and reliance on preprocessing \cite{yu2023deep}, \cite{logrepresentation}, \cite{landauer2023deep}, \cite{pinpoint}. On the other hand, simple rule-based detectors, such as sequence length, have been shown to reach good detection performance on commonly used HDFS dataset in anomaly detection research \cite{criticalreview}. 

In industrial settings, there are often large amounts of data, but little time, which makes simple detectors much more sought after. Additionally, labeled data is rare because the process is labor intensive, so it is better if our detectors can utilize the data in its rawest form. This study addresses several factors when one has to operate with label free datasets.  

Conducting anomaly detection on software logs contains various options that a practitioner can choose from. These options include everything from data selection to model utilization. Studies on the topic generally select certain settings and compare only one selected factor, e.g., model selection \cite{literaturereview}. This makes it difficult to compare results across studies as the other settings of the studies are different. In this study we, aim to conduct empirical evaluation across various factors to examine the impact that individual factors might introduce. As such, this study is also heavily motivated by challenges in practical settings where the practitioner is presented with various options and does not know what to choose. This study aims to provide insights and guidelines on which approaches work well given the characteristics of the data.

Past work in unsupervised log anomaly detection is limited as we only found a handful of papers that give due consideration to preprocessing and data selection. For example, a recent study that aimed to provide a comprehensive view on techniques for log-based anomaly detection admits that the lack of preprocessing optimization can affect detection accuracy and generalizability of results \cite{acomprehensive}. A second major gap in existing work is on the data representations, that is, in what format it is given to the model. Past work has shown that traditional machine learning (ML) models work better with classical log representation techniques, while deep learning models work better with semantic-based representations \cite{logrepresentation}. This study will focus on lightweight representations on lightweight models as we are optimizing for speed. A more thorough summary of previous work is given in Section \ref{sec:background}.

This study extends our previous paper \cite{speedandperf}. The previous paper had 4 datasets; here we introduce 3 new ones. We also include two new preprocessing approaches which results in four times more experiments than previously. In total we have 672 experimental setups which are the result of iterating across six factors (dataset, model, representation, and three preprosessing options). Furthermore, this extension of experimental setups enables us to use generalized linear models that make the evidence in this paper far more robust than what we presented previously. These extensions makes this paper the most comprehensive study regarding anomaly detection on software logs with fast unsupervised methods. We formulate the following research questions.

\begin{itemize}
    \item \textbf{RQ1:} What effect do the model, representation and preprocessing options have on anomaly detection performance on software logs?
    \item \textbf{RQ2:} To what extent is the variability explained by the dataset?
    \item \textbf{RQ3:} How do the computational costs of different models, representations, and preprocessing approaches compare?
\end{itemize}

The distinction between these two research questions is that in RQ1, we address factors that the practitioner might be able to affect, e.g., they could change the model. However, data is what they want to analyze in the first place. Therefore, our dataset-specific analysis in RQ2 will focus on examining the variability and generalizability of the results between datasets in relation to the selection in RQ1. RQ3 aims to address the practical reality of the execution times throughout the anomaly detection pipeline while considering the different factors in this study. 


This study will proceed by providing an overview of the related work regarding unsupervised anomaly detection on software logs in Section \ref{sec:background}. In the Research Methods section (Section 3) we describe the preprocessing, datasets, metrics and models in more detail as well as give an overview of the statistical approaches used in the study. The Results section (Section 4) presents the results by starting from a generic overview and proceeding to more specific findings. The findings along with their implications and key contributions are discussed in the Discussion section (Section 5). Finally, we conclude the study in Section 6.

\section{Background}
\label{sec:background}

In this section, we first introduce an overview taxonomy of prior works in Section \ref{sec:taxonomy}. Then we explore the most notable prior works in more detail in Section \ref{sec:representativeworks}

\subsection{Taxonomy overview}
\label{sec:taxonomy}
To position our work, we present a taxonomy of relevant previous work in Table \ref{tab:taxonomy_prior_work}. The table is categorized by the factors in our empirical evaluation. For a study to be included in the taxonomy, it must be a primary study that evaluates anomaly detection for software logs. Furthermore, it either needs to include an unsupervised model in the analysis \cite{expreport}, \cite{loganomaly}, \cite{autolog}, \cite{withoutparsing}, \cite{howfar}, \cite{tplogad}, \cite{deeplog}, \cite{attgru}, \cite{plelog}, \cite{logfit}, \cite{mlad}, or it needs to do a comprehensive analysis on some of the other studied factors (representations, anomaly free training data, shuffled training data, custom message vs. full row) as in  \cite{logrepresentation}, \cite{criticalreview}, \cite{acomprehensive}, \cite{Wu2025}. 

\begin{table*}
\centering
\caption{Taxonomy of Prior Work by Studied Factors}
\label{tab:taxonomy_prior_work}
\begin{tabular}{ll|c}
\hline
Category & Value & Representative Works \\
\hline
\multirow{7}{*}{Dataset} 
 & BGL & \cite{speedandperf}, \cite{acomprehensive}, \cite{logrepresentation}, \cite{expreport}, \cite{loganomaly}, \cite{autolog}, \cite{withoutparsing}, \cite{howfar}, \\
 & & \cite{criticalreview}, \cite{tplogad}, \cite{deeplog}, \cite{plelog}, \cite{logbert}, \cite{logfit}, \cite{mlad} \\
 & HDFS & \cite{speedandperf}, \cite{acomprehensive}, \cite{logrepresentation}, \cite{expreport}, \cite{loganomaly}, \cite{withoutparsing}, \cite{howfar}, \cite{criticalreview}, \cite{attgru},
 \\
 & & \cite{tplogad}, \cite{deeplog}, \cite{plelog}, \cite{logbert}, \cite{logfit}, \cite{mlad} \\
 & Hadoop & \cite{speedandperf}, \cite{acomprehensive}, \cite{autolog}, \cite{criticalreview} \\
 & Spirit & \cite{logrepresentation}, \cite{acomprehensive}, \cite{withoutparsing}, \cite{howfar}, \cite{criticalreview}, \cite{tplogad} \\
 & Thunderbird & \cite{speedandperf}, \cite{acomprehensive}, \cite{logrepresentation}, \cite{withoutparsing}, \cite{howfar}, \cite{criticalreview}, \cite{tplogad}, \cite{logfit}, \cite{mlad} \\
 & Liberty
 & - \\
 & LO2Token
 & - \\
\hline
\multirow{3}{*}{Method} 
 & Tree-based ML & \cite{speedandperf}, \cite{acomprehensive}, \cite{autolog} \\
 & Cluster-based ML & \cite{speedandperf}, \cite{acomprehensive}, \cite{loganomaly}, \cite{autolog}, \cite{plelog}\\
 & Statistical & \cite{speedandperf}, \cite{acomprehensive}, \cite{expreport}, \cite{loganomaly}, \cite{criticalreview}, \cite{attgru}, \cite{deeplog}, \cite{plelog}, \cite{logbert} \\
\hline
\multirow{3}{*}{Representation} 
 & Words & \cite{speedandperf}, \cite{logrepresentation}, \cite{withoutparsing} \\
 & Trigrams & \cite{speedandperf} \\
 & Parsed event & \cite{speedandperf}, \cite{logrepresentation}, \cite{expreport}, \cite{loganomaly}, \cite{autolog}, \cite{howfar}, \cite{criticalreview}, \cite{attgru}, \cite{tplogad}, \cite{logbert} \\
\hline
\multirow{2}{*}{Training data} 
 & Anomaly free & \cite{speedandperf}, \cite{expreport} \\
 & Chronological vs. random
 & \cite{howfar}, \cite{logrepresentation} \\
\hline
Custom loader & Extract log message
& - \\
\hline
\end{tabular}
\end{table*}

The total list of representative works as noted by their primary tool or author is as follows: Loglizer \cite{expreport}, LogAnomaly \cite{loganomaly}, DeepLog \cite{deeplog}, AutoLog \cite{autolog}, NeuralLog \cite{withoutparsing}, Wu et al. \cite{logrepresentation}, Le and Chang \cite{howfar}, Landauer et al. \cite{criticalreview}, Ali et al. \cite{acomprehensive}, ATT-GRU \cite{attgru}, TPLogAD \cite{tplogad}, PLELog \cite{plelog}, LogFiT \cite{logfit}, and MLAD \cite{mlad}.

The taxonomy illustrates which datasets, methods, representations and preprocessing approaches used in this study have been studied in the past. We can see which approaches have been popular and which have received less attention. For instance, almost all of the studies use the BGL and HDFS datasets. Similarly, almost all studies use a log parser to extract events from the log data. 

The \textit{Dataset} category in the taxonomy contains the datasets used in this study. Some datasets used by other studies are not included here because they use preprocessed IDs which makes using word and trigram representations impossible, e.g., the ADFA dataset \cite{adfa}.  
The Liberty dataset is used in anomaly detection research, and is a fitting dataset for us too, but it is not included in the representative works. LO2Token is a new dataset for this study that is extracted from a recently published dataset \cite{lo2data}. 

The \textit{Method} category contains the models used in the prior works as categorized under broader terms. For instance, this study uses IsolationForest and KMeans models which would go under Tree-based and Cluster-based models respectively. Statistical methods is a broad category that includes both simpler approaches like analyzing frequency distributions and more complex techniques such as PCA. The key distinction is that the statistical models aim to summarize data through closed-form calculations while ML-based processes typically utilize a cost function through an iterative process. Note that for the method section, we included a study as a representative work if it included the appropriate method either as a part of the evaluation or as a novel approach.  

The \textit{Representation} category consists of words, trigrams and parsed event categories. A widely used technique for log representation is to use a clustering algorithm, also known as log parsing, to find out the static part of a log message called the template and use that (or its ID) to represent the logline \cite{literaturereview}. Drain \cite{drain} is one such parser commonly used in software log anomaly detection studies. However, recent studies have demonstrated that there necessarily is not a strong correlation between parsing and anomaly detection accuracy \cite{parserimpact}, and some studies have begun to implement parserless approaches to achieve competitive results, especially when accounting for performance \cite{Sun2026}, \cite{speedandperf}, \cite{withoutparsing}. Typically parserless representations split the log message by whitespace and apply some simple regex or other normalization. The use of large language models (LLMs) for parsing has also been studied as part of semi-supervised anomaly detection pipeline in the SemiRALD approach \cite{semirald}. It achieved better parsing accuracy than common parsers with competitive execution time despite the complex nature of LLMs. However, LLMs are not currently scalable for large amounts of completely unsupervised data. It should be noted that comparing execution times between studies is difficult due to the vastly different environments. For instance, \cite{semirald} reported that parsing HDFS with Drain takes over 100 minutes, while another recent study parsed 1 million lines (21\%) in under two minutes \cite{XDrain} with the same parser and dataset. 

The \textit{Training data} category examines two options for training data. In short, the first considers whether the training data includes anomalies or are they filtered out. If there are anomalies in the training data, an unsupervised model has no way of distinguishing faulty behavior from simply rare behavior. It is still worth studying both options, as some datasets might show good results when effectively equating anomalousness with rarity \cite{speedandperf}. The second option for training data considers the impact of using chronological versus random training data selection. Using completely random training data selection leads to some training data points being chronologically later than some test data points. This can be considered as a form of data leakage \cite{howfar}. However, in a practical setting it can be possible to isolate the anomalous region chronologically, which is why we study the effect of it in this study. 

The final option examines the effect of using a \textit{custom loader} for the dataset to only extract the message from the log line (i.e., discarding the metadata like timestamps). We found no studies that explicitly study this with regards to unsupervised anomaly detection, yet it has become the common practice to extract the log message from the log line. We expect that the metadata is irrelevant noise most of the time, but at the same time, if the models are able to handle it gracefully, the metadata can help identify genuine anomalies.   

Regarding the training data category and custom loader options, we included a study in the taxonomy if it studied as opposed to just utilized a given approach. For instance, if the study uses training data where the anomalies are filtered out, but does not study the effect of filtering, it is not included in the taxonomy. This is done because papers do not often report the preprocessing settings so the results would be arbitrarily determined by the authors decision to report something rather than what they actually did. This also explains why there are only a few studies in the representative works for the preprocessing approaches.  

\subsection{Representative works}
\label{sec:representativeworks}
As this paper directly follows our work done in \cite{speedandperf}, it is naturally the most prevalent representative work in the taxonomy. It studied anomaly detection performance of four unsupervised anomaly detection models on five different datasets. It also studied the effect of data representation, which refers to the format in which the data is given to the model. In addition, it studied the impact of filtering out anomalies from the training data. Although the approach requires training data for only one class (i.e. normal), the models themselves work on purely unsupervised principles. This is also referred to as novelty detection \cite{language_novelty}. The assumption in this setup is that the practitioner would have access to clean (i.e., anomaly free) logs which they would use as training data to gain better performance on unsupervised models \cite{speedandperf}. 

Crucially, this paper will improve upon our previous one by including more datasets and studying more preprocessing approaches. One such approach involves whether or not the data is shuffled before being assigned to training and test sets. So far, there have been only a handful of studies assessing the effect of chronological vs. shuffled training data \cite{criticalreview}, and in many cases the selection is not reported in the study. The third preprocessing option is whether or not to use a custom loader to extract only the log message from the log line before performing any other preprocessing. As explained in the previous section, we found that the research related to using a custom loader for unsupervised anomaly detection is very limited.

One of the most prevalent paper in the taxonomy \cite{expreport} used a tool called Loglizer. When the paper was published in 2021, they reported that Loglizer\footnote{https://github.com/logpai/loglizer} is the only available toolkit for log anomaly analysis. 
The study that reported results with the Loglizer \cite{expreport} toolkit used several methods, of which Invariant Mining achieved the best results with an F1 score of 0.91 on both HDFS and BGL datasets. The authors note the challenges of setting and testing the anomaly detection threshold in unsupervised models \cite{expreport}. The unsupevised models used in that study were also much more time consuming. For example, 1.5 million lines on HDFS took more than 1000 seconds with Invariant Mining. Unlike most other papers, the study also considered the amount of anomalies in the training data by dynamically adjusting it \cite{expreport}. This is noted as representing the "anomaly free training data" category in the taxonomy. 


LogAnomaly \cite{loganomaly} is an approach that utilizes an unsupervised LSTM model that combines word and template vectors and manages to achieve F1 scores of 0.96 (BGL) and 0.95 (HDFS). The study provided comparison also to lighter unsupervised machine learning and statistical models. These include PCA and clustering models. While the method reached good results, other studies have shown that deep learning-based models are sensitive to hyperparameter tuning and computationally expensive \cite{empiricalml}. 

AutoLog \cite{autolog} is another paper that presents a new deep learning based approach. It focuses on training the model with anomaly free data and predicting anomalies based on reconstruction error. Their model uses a deep autoencoder and manages to reach F1-scores of 0.95 on BGL and 0.97 on HDFS. Additionally, they compare their model with unsupervised models such as the IsolationForest which reached worse scores of 0.70 on BGL and 0.61 on HDFS. As the authors utilize anomaly free training data, the experimental approach should be similar to the anomaly free training data setup used in this study. 


A study \cite{logrepresentation} conducted a thorough analysis on log representations used in anomaly detection. The authors applied several classical and semantic-based representations on traditional and deep-learning models. One of the key findings was that the traditional models performed better with classical representations, which is a major motivation for our work. The study did not include unsupervised methods but it was included in the taxonomy because it provides the most thorough analysis on log representations. Regarding their preprocessing philosophy they explicitly state only using randomly split training data without consider the chronological order of events \cite{logrepresentation}.

So far, there have been few studies on the impact of using chronologically sorted vs. randomly selected training data. A recent study that critically evaluated common datasets included it as a feature to study \cite{howfar}. They found that on some scenarios there is a significant improvement in performance when using randomly selected data, as it provides a more diverse set of samples. However, model selection played an important role, as supervised models did not experience a similar performance boost on randomly selected training data.

Some recent studies have shown increased performance over classic approaches like PCA and Invariant Mining by using attention, temporal neural networks, and richer log representations. NeuralLog \cite{withoutparsing} introduces an anomaly detection approach that works without parsing and as such is one of the few studies in the taxonomy that does not rely on parsed events. By doing this, it aims to solve the issue of Out-of-vocabulary (OOV) words and semantic misunderstandings. NeuralLog achieves great results on several datasets, but employs a BERT encoder and transformer-based classification \cite{withoutparsing}, which are relatively time-consuming and resource intensive approaches. ATT-GRU \cite{attgru} adds attention to a GRU model and beats DeepLog, PCA, and IM with higher accuracy and lower latency, but its complex design and limited implementation details make it hard to judge whether the extra complexity is worthwhile. It depends on labeled historical data, which is often unrealistic in mostly unsupervised log analysis settings. TPLogAD \cite{tplogad} combines BERT-based template and parameter embeddings with an attention-enhanced BiLSTM in a fully unsupervised framework, achieving strong results against several state-of-the-art baselines, but at the cost of a heavy, intricate architecture and unclear procedures that may not scale well to real-time, large production systems.

Overall, these works show a consistent pattern. More complex neural models tend to perform better to a certain extent, but they introduce higher computational cost, complex tuning, and reduced transparency and reproducibility. This is despite the fact that the recent studies have highlighted the simplicity of studied public datasets \cite{Wu2025} and that classic ML can reach competitive results \cite{yu2023deep}. The practical concerns regarding the benchmark studies raise questions whether the approaches can provide robust, and cost-effective solutions for large-scale, log anomaly detection in real-world environments.

\section{Research methods}
This section describes the research methods as divided into three sections. In the first section, we explain the details regarding the experimental design. The design is split into subsections that match the structure of the taxonomy used to position our work in Section \ref{sec:taxonomy} as well as the experimental procedure. Next, we describe the metrics used in anomaly detection and the reasoning for selecting those metrics. Finally, we outline details about the technical implementation of the study.

\subsection{Experimental design}

The four categories of the experimental design that consider the factors being studied are datasets, models, representations, and preprocessing options. This section describes each in detail. Additionally, we explain details about the experimental procedures that we will use to conduct the experiments. 

\subsubsection{Datasets}
\label{sec:datasets}
The datasets used in this study are BGL \cite{bgl}, Thunderbird \cite{bgl}, HDFS \cite{hdfs}, Hadoop \cite{hadoop}, Spirit \cite{bgl}, Liberty \cite{bgl}, and LO2Token. BGL, HDFS, and Hadoop are accessible via LogHub \cite{loghub}. Thunderbird, Spirit, and Liberty can be downloaded from the Usenix website \footnote{\url{https://www.usenix.org/cfdr-data\#hpc4}}. LO2Token dataset is included in the replication package of this study in Zenodo \cite{replication}. Table~\ref{tab:datasummaries} shows the summary of the datasets as split by event-based and sequence-based datasets.

\begin{table*}[t]
\caption{Dataset Summaries}
\label{tab:datasummaries}

\vspace{0.5em}
\textbf{(a) Event-based Datasets}
\begin{tabular}{lllll}
\hline
 & \textbf{BGL} & \textbf{Thunderbird} & \textbf{Spirit} & \textbf{Liberty} \\
\hline
Log lines & 4,747,963 & 211,212,174 & 272,298,895 & 265,569,119 \\
Normal lines & 4,399,503 & 193,744,715 & 90,656,198 & 168,478,341 \\
Anomalous lines & 348,460 & 17,467,459 & 181,642,697 & 97,090,778 \\
Data used  & 100\%  & 3\% & 2\% & 2\% \\
Train portion  & 5\% & 5\% & 5\% & 5\% \\
\hline
\end{tabular}

\vspace{1em}
\textbf{(b) Sequence-based Datasets}
\begin{tabular}{llll}
\hline
 & \textbf{HDFS} & \textbf{Hadoop} & \textbf{LO2Token} \\
\hline
Log lines & 11,175,629 & 394,308 & 191,850,132 \\
Normal sequences & 558,223 & 11 & 1,727 \\
Anomalous sequences & 16,838 & 43 & 91,531 \\
Data used & 100\% & 100\% & 1.8\%* \\
Train portion & 5\% & 50\% & 5\% \\
\hline
\end{tabular}
\end{table*}

The distinction between event- and sequence-based datasets is important as they use a fundamentally different approach for anomaly detection. The event-based datasets have labels on each logline, and as such we can use the anomaly detection models to score each line and assess that accurately. The sequence-based datasets have labels on sequence level which means that normal lines within that sequence would also be labeled as anomalies in the event-based approach. Therefore, the sequence-based approach works by appending the contents of the whole sequence into a single prediction. The most significant difference with the sequence and event based datasets for anomaly detection is that when using event ID as the log representation, the sequence based datasets have multiple IDs to influence the prediction while event based ones only have one. For the word and trigram representations, the individual items used in the model just become much longer but they work effectively the same way.

In general, logs have relatively homogeneous content which implies that a small percentage for model training training is enough. DeepLog demonstrated that for HDFS a less than 1\% split for training is sufficient \cite{deeplog}. As was done in the previous paper \cite{speedandperf}, this study uses 5\% of the data (after reduction) for training for every dataset except Hadoop which uses 50\% due to a very small number of sequences. For the large event-based dataset, we only utilize a small percentage of data to be able to run them and have their size comparable to BGL. The values are on Table \ref{tab:datasummaries}. For the sequence based datasets, we utilize the whole data except for LO2Token which selects a random error case along with the correct test. This results in a balanced dataset with 1727 normal and 1727 anomalous sequences, and effectively reduces the data usage to approximately 1.8\% of the total data. This is indicated by the asterisk (*) in Table \ref{tab:datasummaries}.

LO2Token is the new dataset created for this study based on the recently published LO2 microservice dataset \cite{lo2data}. The LO2 dataset was created by running 53 negative test setups along with a positive test for complete 1740 runs. Because it is a microservice dataset, each seven services have their own log files. Normalizing anomaly detection scoring between different datatypes has proven to be difficult and requires purpose-built solutions \cite{myllari2025ladle}. Such multi-service analysis is not in the scope of this study, so for the LO2Token dataset we only selected the logs from the Token service. After removing runs with issues there was a total of 1727 runs where each have the 53 negative tests with errors and the normal positive test. This results in a total of 93,258 log files which are effectively treated as sequences in this study. For the LO2Token dataset correct sequences are reduced by a factor = 12.744 to make their size not significant as they were ran for a longer time.

We chose Token as the included service because the initial dataset paper \cite{lo2data} showed that it appears to be the best service to predict the prevalence of an error with an overall average F1 score of 0.980 when using the supervised Decision Tree model. Initial tests quickly showed that unsupervised models do not reach such performance and the results are more in line with the other commonly used datasets in anomaly detection research.

\subsubsection{Models}
\label{sec:models}

For this study we used the same model selection as the previous work \cite{speedandperf} . During our initial testing, we evaluated far wider range of unsupervised models, but some popular unsupervised approaches—such as Local Outlier Factor and One-class SVM—proved computationally too slow for our needs and unlikely to therefore be used with large industrial data sets. 

The models include two established models in the Sklearn library (Isolation Forest and KMeans) and two simpler custom models we developed to have more performance efficient alternatives (Out-of-vocabulary Detector and Rarity Model). All of the models work by producing a numeric value that correspond to how anomalous a given event or sequence is. As the models work identically with all three log representations (words, trigrams, event IDs), in this section we refer to them all collectively as "terms".

One model implemented with Sklearn \cite{scikit-learn} is \textbf{Isolation Forest (IF)}. It is based on random isolation of features and splitting between maximum and minimum values of selected feature \cite{scikit-learn}. Due to their nature, anomalies are easier to separate than normal data. Based on that, it can be determined that the number of partitions it takes to isolate a data point corresponds with how likely it is to be normal. 

The second model implemented with Sklearn is the \textbf{KMeans} clustering algorithm \cite{scikit-learn}. It works by starting with random centroids and assigning data points to the closest one. As data is added, the centroids are updated based on the mean of all the data points in that centroid. The trained model should have clearly defined clusters around the centroids. For prediction, we give the anomaly score as the distance to the nearest centroid. 

We started to design the \textbf{Out-of-vocabulary Detector (OOVD)} as a tool for aiding other models. However, when training the model with only normal data OOV terms are great candidates for being anomalies as the task of anomaly detection turns into novelty detection \cite{language_novelty}.  As such the OOVD by itself can produce good results when there is clean training data available. 

It is important to note that the OOVD can not work if there are anomalies in the training data vocabulary. Other models can work with the anomalous data because they determine the anomalousness of a term is by its relation to other terms, but OOVD only checks whether it belongs to a set. Therefore, if the set is created randomly the results will be random as well. 

Counting OOV terms in millions of lists individually would be a slow operation so we utilize a strategy that takes advantage of the way OOV terms are handled by vectorizers. When transforming the test data with CountVectorizer that was fitted with training data, the OOV terms will automatically get the value 0. Then we can simply subtract the number of in-vocabulary terms calculated by the CountVectorizer from the total number of terms to get the number of OOV terms. Working with total counts as such can be hundreds of times faster than matching the words individually in a for-loop.

We implemented a simple class based on negative logarithm that we called the \textbf{Rarity Model (RM)}. In essence, negative logarithm measures the relative infrequency of individual terms in the data that increases based on their rarity. While utilizing negative logarithm in such a way is not novel, our aim was to create a very efficient baseline model. For implementation, the rarity scores are first precalculated for each term in the training data and assigned into a vector. Next, the dot product of the precalculated vector and the sparse matrix generated by the Tf-idf vectorizer from the test data is calculated. The total rarity score for the log message will be the dot product divided by the total number of terms in the message. Given its simple design and reliance on pre-built matrix multiplication routines, the RM functions with remarkable speed \cite{speedandperf}.

\subsubsection{Representations}
\label{sec:representations}

In this study, creating representations means that the normalized logs are transformed into event IDs or lists consisting of trigrams or words. Words are extracted simply by using whitespace as the separator. Event IDs are created with the Drain parser \cite{drain}. 
By "trigrams", we refer to character trigrams, that is, three consecutive characters in the log message. For example, the first three trigrams of the message "Stopping..." would be "Sto", "top" and "opp".  
An example of these representations along with their raw and normalized log messages is shown in Table~\ref{tab:preprocess}.

\begin{table*}[ht]
\caption{Example message from the preprocessed BGL dataset}
\label{tab:preprocess}
\centering
\begin{tabular}{|p{0.15\textwidth}|p{0.2\textwidth}|p{0.2\textwidth}|p{0.2\textwidth}|p{0.1\textwidth}|}
\cline{3-5}
\multicolumn{2}{c|}{} & \multicolumn{3}{c|}{\textbf{Representations}} \\ 
\hline
\textbf{Raw log message} & \textbf{Normalized message} & \textbf{Words} & \textbf{Trigrams} & \textbf{Event ID} \\ 
\hline
'4 ddr error(s) detected and corrected on rank 0, symbol 11 over 20609 seconds' & '0 ddr error(s) detected and corrected on rank 0, symbol 0 over 0 seconds' & [ '0', 'ddr', 'error(s)', 'detected', 'and', 'corrected', 'on', 'rank', '0', 'symbol', '0', 'over', '0', 'seconds'] & ['0 d', 'dd', 'ddr', 'dr ', 'r e', 'er ', 'err', 'rro', 'ror', 'or(', '(r(s', '(s)', 's) ', ') d', ' d ', 'de', 'det', 'ete', 'tec', 'ect', ...] & 'e10' \\ 
\hline
\end{tabular}
\end{table*}

\subsubsection{Preprocessing options}
\label{sec:preprocessingoptions}
In the study, we use three preprocessing options that can be set to either true or false. First one is the \textit{Filter} option which, when set to true, will remove the anomalous entries from the training data. We acknowledge that in an industrial setting, it is not feasible to filter out anomalies. However, the industry can separate logs that are gathered during the system's normal operation that should work as the training data. Hence, the \textit{anomaly free} setup does require a certain level of curation from the log owner but at the a very high level. 

Using the anomaly data changes fundamentally what is being examined. When there are anomalies in the training data but no labels, it is impossible to identify which events are true anomalies and which are not. Then the anomaly detection essentially relies on the assumption that rarity equates anomalousness. In contrast, with the anomaly free setup, anomalousness is instead based on the deviation from normal. 

The second option is \textit{Shuffle} which shuffles the data before assigning training and test sets. If the data is not shuffled, the data will be in chronological order so that all training data is earlier than all test data.  

The \textit{Custom} option refers to whether or not we use a custom loader for the data that retrieves only the message part from the log line. In contrast, a generic loader uses the whole row. For the event-based datasets, we use a generic "Raw Loader" which simply retrieves each line as their own event. Although, if it contains labels, we remove it afterwards. For the sequence based datasets we can not use a completely generic loader as we need to create the sequences. In addition, multi-line entries (e.g., stacktraces) are combined for practical purposes. Within a logline, however, there is no custom logic. Table \ref{tab:log_parsing} shows a sample from the HDFS dataset as split into separate fields. When using a custom loader, only the log message shown in cursive is selected for analysis.

\begin{table*}[h]
\centering
\caption{Example of field extraction from HDFS logs using custom loader}
\label{tab:log_parsing}
\begin{tabular}{p{0.12\textwidth} p{0.07\textwidth} p{0.2\textwidth} p{0.45\textwidth}}
\toprule
\textbf{Timestamp} & \textbf{Level} & \textbf{Component} & \textbf{Message} \\
\midrule
081110 002221 4855 & INFO & dfs.DataNode\$ PacketResponder & \textit{PacketResponder 0 for block blk\_3259872359515191972 terminating} \\
081110 002223 4824 & INFO & dfs.DataNode\$ PacketResponder & \textit{PacketResponder 1 for block blk\_5963816136267066159 terminating} \\
081110 002253 34   & INFO & dfs.FSNamesystem & \textit{BLOCK* NameSystem.allocateBlock: /user/root/sortrand/\_temporary/part-00138. blk\_-210021574616486609} \\
081110 002337 13   & INFO & dfs.DataBlockScanner & \textit{Verification succeeded for blk\_-1547954353065580372} \\
\bottomrule
\end{tabular}
\end{table*}

The rationale for studying the impact of using a custom loader is that it takes effort to create it while the benefit remains unclear. On one hand, the general idea is that the anomalous behavior should appear in the log message, so that would be the only relevant field in the log line. Furthermore, log parsers typically only utilize the log message part. On the other hand, excluding the other fields removes information that might possibly be better indicators of the labeled anomalies than the log message itself.  

We will study all eight combinations of the three preprocessing options. Table \ref{tab:fsc} provides textual descriptions on each setup, but further in the paper we will refer to them with abbreviations. Specifically, F for Filtered (or Anomaly Free), S for Shuffle, and C for Custom Loader. For the eight combinations, we use a string identifier that denotes whether specific preprocessing option is on or off.

\begin{table*}[h]
\caption{Comparison of different experimental setups and their effects on data processing}
\label{tab:fsc}
\centering
\begin{tabular}{lcccp{10cm}}
\hline
\textbf{Setup name} & \textbf{F} & \textbf{S} & \textbf{C} & \textbf{Description} \\
\hline
f0\_s0\_c0 & & & & Uses full log rows. Data split chronologically into train/test. Anomalies retained in training data. \\
\hline
f1\_s0\_c0 & \checkmark & & & Uses full log rows. Data split chronologically into train/test. Anomalies removed from training data. \\
\hline
f0\_s1\_c0 & & \checkmark & & Uses full log rows. Data shuffled before train/test split. Anomalies retained in training data. \\
\hline
f0\_s0\_c1 & & & \checkmark & Loader selects log message only. Data split chronologically into train/test. Anomalies retained in training data. \\
\hline
f1\_s1\_c0 & \checkmark & \checkmark & & Uses full log rows. Data shuffled before train/test split. Anomalies removed from training data. \\
\hline
f1\_s0\_c1 & \checkmark & & \checkmark & Loader selects log message only. Data split chronologically into train/test. Anomalies removed from training data. \\
\hline
f0\_s1\_c1 & & \checkmark & \checkmark & Loader selects log message only. Data shuffled before train/test split. Anomalies retained in training data. \\
\hline
f1\_s1\_c1 & \checkmark & \checkmark & \checkmark & Loader selects log message only. Data shuffled before train/test split. Anomalies removed from training data. \\
\hline
\end{tabular}
\label{tab:setup_comparison}
\end{table*}

\subsubsection{Experimental Procedure and Statistical Evaluation}

To conduct the study we will run all the experimental combinations five times similar to \cite{expreport}. The combinations include 7 datasets, 4 models, 3 representations, 8 preprocessing option setups, and 5 repeats which result in a total of 3360 individual data points for both metrics. We will use statistical approaches to evaluate these results after conducting all experiments. In particular, we utilize a generalized linear model (GLM). We will first fit the model by using the variables independently to get a general overview of the coefficients. After that we will apply the dataset as an interaction term on each different variable to be effectively be able to assess the impact of the dataset on that variable. E.g., if some anomaly detection model works well on some dataset. Due to the limitations of using F1-score (as explained in the next section, \ref{sec:metrics}) and the scope of the paper, we will conduct the GLM analysis only using the AUCROC metric.  

After the GLM with interaction terms is fitted, we aim to do further analysis on the residuals which means the difference of the actual value and the value predicted by the GLM. The residuals can highlight scenarios where the model does not perform well, which in practice means that there are some results that are not explained by the individual variables or their interactions with the dataset. These could be, for example, interactions between the model and representation selection. 

Due to the exploratory nature of the study, we aim to conduct additional experiments after the regression analysis to gain further insights regarding the most important findings. The exact details of these experiments depend on the type of finding and are elaborated at the end of the Result section (\ref{sec:supplementary}).

\subsection{Anomaly Detection Metrics}
\label{sec:metrics}

In the study we will analyze model performance primarily with AUC-ROC, but we will provide an overview of the F1-scores too. AUC-ROC \cite{aucroc} is a combination of the abbreviations Area Under Curve (AUC) and Receiver Operating Characteristic (ROC). ROC curve is a plot of the true positive rate against the false positive rate across a range of thresholds. With AUC we refer to the area under this curve. This area goes from 0 to 1 and in practice it means the probability that a randomly selected positive sample is ranked higher than a negative one \cite{aucroc}. F1-score is the most common metric in anomaly detection that measures the harmonic mean of precision and recall.

The main reason for emphasizing AUC-ROC in this study is that it is not dependent on a threshold. With supervised models you could assess the model specific anomaly scores in relation to the labels and adjust the threshold accordingly. In unsupervised models this is not possible which leads to subjective threshold selection.

As realistic threshold selection (i.e., without labels) for the F1-score is out of scope of this paper, we will utilize labels in a Bayesian optimization loop that finds the best threshold over a maximum of 20 iterations. We emphasize that this is not possible without labels, and as such, the F1-scores serve as an example of what could be possible. 


\subsection{Implementation}

\subsubsection{Computing environment}

All the tests are run on a PC with Intel Core i7-13700K, which has 16 cores at 4.4 GHz for P-Cores and 3.4 GHz for E-Cores. Regarding the computational performance, all experimental setups are limited by the CPU only with some operations using only a single core, while others can be distributed to multiple cores. Some of the setups require a large amount of memory. Our PC has 128 GB of physical RAM and uses 128 GB of SSD storage as virtual memory. For running the experiments, we utilize a modified version of the LogLead\footnote{https://github.com/EvoTestOps/LogLead} tool that was designed to benchmark anomaly detection algorithms and log representations. The replication package is publicly available on Zenodo\footnote{Ready but not published.}. 

\subsubsection{The anomaly detection pipeline}
The anomaly detection pipeline starts from preprocessing steps that are common for all experimental setups. For these we follow exactly the same process as in the previous study \cite{speedandperf}. First, the data is loaded directly from raw log files to Polars \cite{Vink2023Polars} dataframes.
Next, the logs are normalized to reduce the effect of parameters (non-static parts of the message). However, due to placing strong emphasis on performance time we only perform three operations with fast Polars-native string operations: 
\begin{enumerate}
    \item Convert all letters to lowercase, 
    \item Replace every numeral with the digit '0',
    \item Reduce any consecutive series of zeros to a single '0'.
\end{enumerate}
For example, the string "Time 12:34:56" would be converted to "time 0:0:0". When it comes to words and trigrams, normalization will make the resulting vocabulary smaller. Although the primary reason for this is to improve the accuracy due to minimizing the effect of parameters, a smaller vocabulary also makes the model faster. 

In the next step, representations are created as explained in the section \ref{sec:representations}. The representations are then vectorized so that they can be used with the machine learning models. For most models, we use the TfidfVectorizer \cite{scikit-learn}, which is a common technique for feature extraction that adds weight to the most important tokens. However, for the Out-of-vocabulary Detector we utilize absolute counts so we use the CountVectorizer. 
 
After preprocessing, the data is split into training and test sets as defined in section \ref{sec:datasets}. Finally, actual training and prediction is performed as defined in the Sklearn library for KMeans and IF, while OOVD and RM use similar custom functions. In addition to binary prediction, all the models are also able to produce a score that corresponds to the perceived anomalousness. In fact, the models work by introducing a threshold on those scores. For the analysis of the results, the scores are crucial as it enables the AUC-ROC analysis.

\section{Results}
\label{sec:results}

This section begins by giving a high level overview of the results. This is followed by regression analysis for anomaly detection performance. Then, we present the results regarding execution time. Finally, we conduct a supplementary experiment to gain insight on the impact of shuffling training data by comparing the unsupervised models to a supervised one. 

\subsection{Overview}

To start, we provide an easy summary of each variable by showing the mean AUCROC and F1-score in Table \ref{tab:performance_metrics}. To calculate the means and standard deviations, we include a single value within a category, but every value from other categories in the pool. The total number is denoted as \emph{n} in the table. For example, the AUCROC mean of 0.658 for BGL includes the results of all models, representations, and Filter/Shuffle/Custom options, as long as the experiment used BGL as the dataset. 

The table suggests that, as expected, the most impactful individual category is the dataset because it shows the widest range means. For instance, the F1 score range for datasets is from 0.276 to 0.884 while the second largest range is in Filtering which improves the F1 score from 0.557 to 0.653. Because the standard deviations are so large and the means are relatively consistent, this table only gives a crude overview. In other words, there are good and bad situations for each parameter, but they cannot be distinguished from such a broad generalization. 

\begin{table*}
\centering
\caption{Result Summary}
\label{tab:performance_metrics}
\begin{tabular}{ll|c|cc|cc}
\hline
Category & Value & n & \multicolumn{2}{c|}{AUCROC} & \multicolumn{2}{c}{F1 Score} \\
 & & & Mean & Std & Mean & Std \\
\hline
\multirow{7}{*}{Dataset} & BGL & 96 & 0.658 & 0.169 & 0.276 & 0.173 \\
 & Hadoop & 96 & 0.645 & 0.179 & 0.884 & 0.120 \\
 & HDFS & 96 & 0.845 & 0.182 & 0.569 & 0.267 \\
 & Liberty & 96 & 0.571 & 0.290 & 0.661 & 0.219 \\
 & LO2Token & 96 & 0.616 & 0.168 & 0.698 & 0.071 \\
 & Spirit & 96 & 0.466 & 0.363 & 0.809 & 0.207 \\
 & Thunderbird & 96 & 0.538 & 0.288 & 0.329 & 0.257 \\
\hline
\multirow{4}{*}{Model} & IsolationForest & 168 & 0.573 & 0.273 & 0.589 & 0.263 \\
 & KMeans & 168 & 0.649 & 0.304 & 0.616 & 0.281 \\
 & RarityModel & 168 & 0.598 & 0.281 & 0.642 & 0.288 \\
 & OOV Detector & 168 & 0.662 & 0.194 & 0.573 & 0.324 \\
\hline
\multirow{3}{*}{Representation} & Words & 224 & 0.656 & 0.272 & 0.630 & 0.281 \\
 & Trigrams & 224 & 0.656 & 0.248 & 0.625 & 0.277 \\
 & Event ID & 224 & 0.549 & 0.272 & 0.560 & 0.309 \\
\hline
\multirow{2}{*}{Anomaly Free} & 0 & 336 & 0.538 & 0.247 & 0.557 & 0.291 \\
 & 1 & 336 & 0.702 & 0.264 & 0.653 & 0.283 \\
\hline
\multirow{2}{*}{Shuffle} & 0 & 336 & 0.625 & 0.255 & 0.598 & 0.289 \\
 & 1 & 336 & 0.617 & 0.282 & 0.611 & 0.293 \\
\hline
\multirow{2}{*}{Custom} & 0 & 336 & 0.623 & 0.274 & 0.607 & 0.291 \\
 & 1 & 336 & 0.618 & 0.263 & 0.603 & 0.291 \\
\hline
\end{tabular}
\end{table*}

\begin{figure*}
    \centering
    \includegraphics[width=0.72\linewidth]{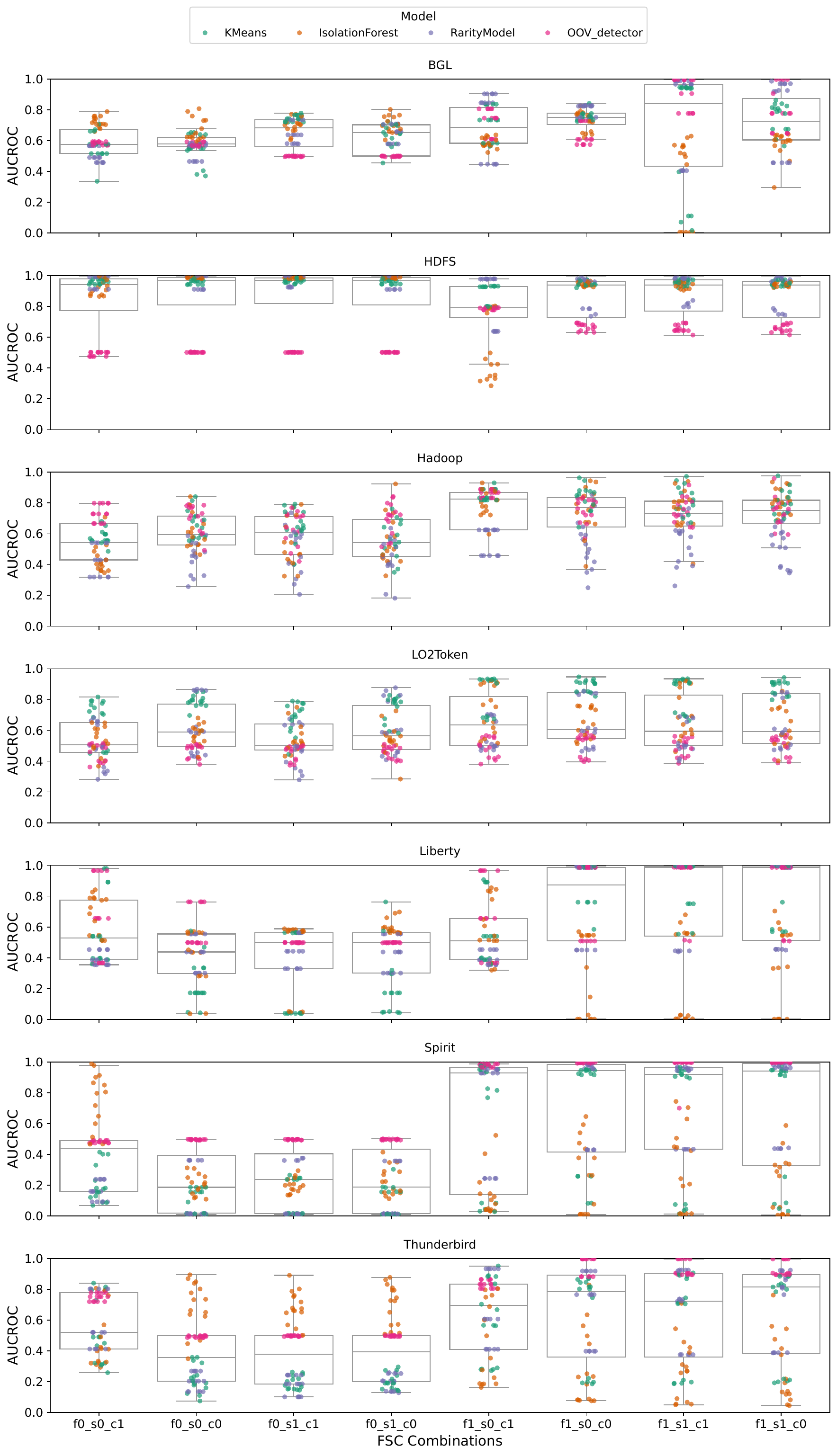}
    \caption{Boxplots by dataset}
    \label{fig:boxplots}
\end{figure*}

Figure \ref{fig:boxplots} shows an overview of all the AUC-ROC results when separated by dataset. There are seven subplots, each representing a dataset, and each plot has eight setups regarding the preprocessing options. Furthermore, individual runs are overlayed on the boxplots with color codes corresponding to model selection. Representation selection can not be distinguished from the plot. To help reading the plot across various setups, consider them in groups. The group of four on the left correspond to anomalous training data while the group of four on the right use anomaly free training data. Groups of two plots alternate between using and not using shuffled data for training. Each plot alternates regarding the use of a custom loader.

The group of datapoints exactly at AUCROC = 0.5 for OOVD is expected to occur when filtering is off. This is due to the nature of OOVD as explained in the OOVD part of section \ref{sec:models}. Furthermore, regarding filtering, when the ratio of anomalous lines to total lines in a dataset is high, we expect the results with anomalous training data to be significantly lower, as the unsupervised models cannot differentiate between common messages that are and are not labeled as anomalies. This raises the question that should they be called anomalies in the first place.

Next we present the key findings for each dataset.

\textbf{BGL} is an average-level dataset in terms of difficulty. The deviation across setups is relatively small. The setup \textit{f1\_s1\_c1} provides the best overall results but also the largest standard deviation with some outliers getting a score of 0. We investigated these outliers and found that they occur with the event ID representation, suggesting that there is an issue with the interaction of the loader and the parser. 

\textbf{HDFS} is the easiest dataset set where the rarity of the log content strongly corresponds to the labeled anomalousness. This can be determined from the fact that almost all models perform well even when there are anomalies in the training data. OOVD performs worse than the other models. On \textit{f1\_s0\_c1} the performance of IsolationForest and RarityModel deteriorate while OOVD improves. This suggests that the number of OOV terms increases and the other models are unable to handle them. It is a reasonable assumption as filtering, not shuffling, and using a custom loader all lead to a smaller vocabulary versus their counterparts. 

\textbf{Hadoop} is the smallest dataset which shows a large variation even on exactly the same setup. The best performance requires anomaly free data and a KMeans or IF model.

\textbf{LO2Token} scores have the smallest standard deviation across all studied datasets. KMeans is generally the best model, while OOVD has bad performance regardless of other factors. We examined the data and found that the log lines that are characteristic to a sequence being labeled as an anomaly also appear in the normal runs, but in much smaller numbers. Regardless, it means that those lines are not considered anomalous by OOVD. The results with the custom loader are worse, which implies that there is relevant information in the discarded part of the logline.

\textbf{Liberty} is completely dependent on filtering, but even then, model selection matters. The low scores of IsolationForest where OOVD gets near perfect AUCROC suggest, again, that there are many OOV terms that the model is unable to handle. Interestingly, \textit{f1\_s0\_c1} has an uniquely low average score among other setups with filtering. In practice, this means that models require either information from the log line (not using custom loader) or in other parts of the dataset (shuffling) to get great performance, but if they get neither, they do not work well. 

\textbf{Spirit} has results quite similar to Liberty. Across all datasets, Spirit has the worst results without filtering due to the high number of anomalous log lines. Essentially, because there are more events labeled as anomalies than normal, the models generally get a worse than random result when using anomalous training data. IsolationForest on \textit{f1\_s0\_c1} is an exception, but that is likely dependent on specific details of the data rather than a generalizable result. 

\textbf{Thunderbird} is similar to the other supercomputer datasets (BGL, Liberty, Spirit, Thunderbird). In terms of difficulty and deviation, it could be characterized as being between BGL and Spirit. Unlike on Spirit, for Thunderbird IsolationForest performs well on exactly the opposite anomaly data.

\subsection{GLM analysis}
\label{sec:regressionanalysis}

To evaluate the effects of each parameter and setting, we employed the generalized linear model (GLM). Linear models require a reference variable which affects the coefficients but not the model itself in terms of predictive capabilities. Based on previous results and ease of interpretation we chose the following reference variables:
\begin{itemize}
    \item Dataset: BGL
    \item Model: KMeans
    \item Representation: Trigrams
    \item Filter: False
    \item Shuffle: False
    \item Custom: False
\end{itemize}

\begin{figure*}[h]
    \centering
    \includegraphics[width=1\linewidth]{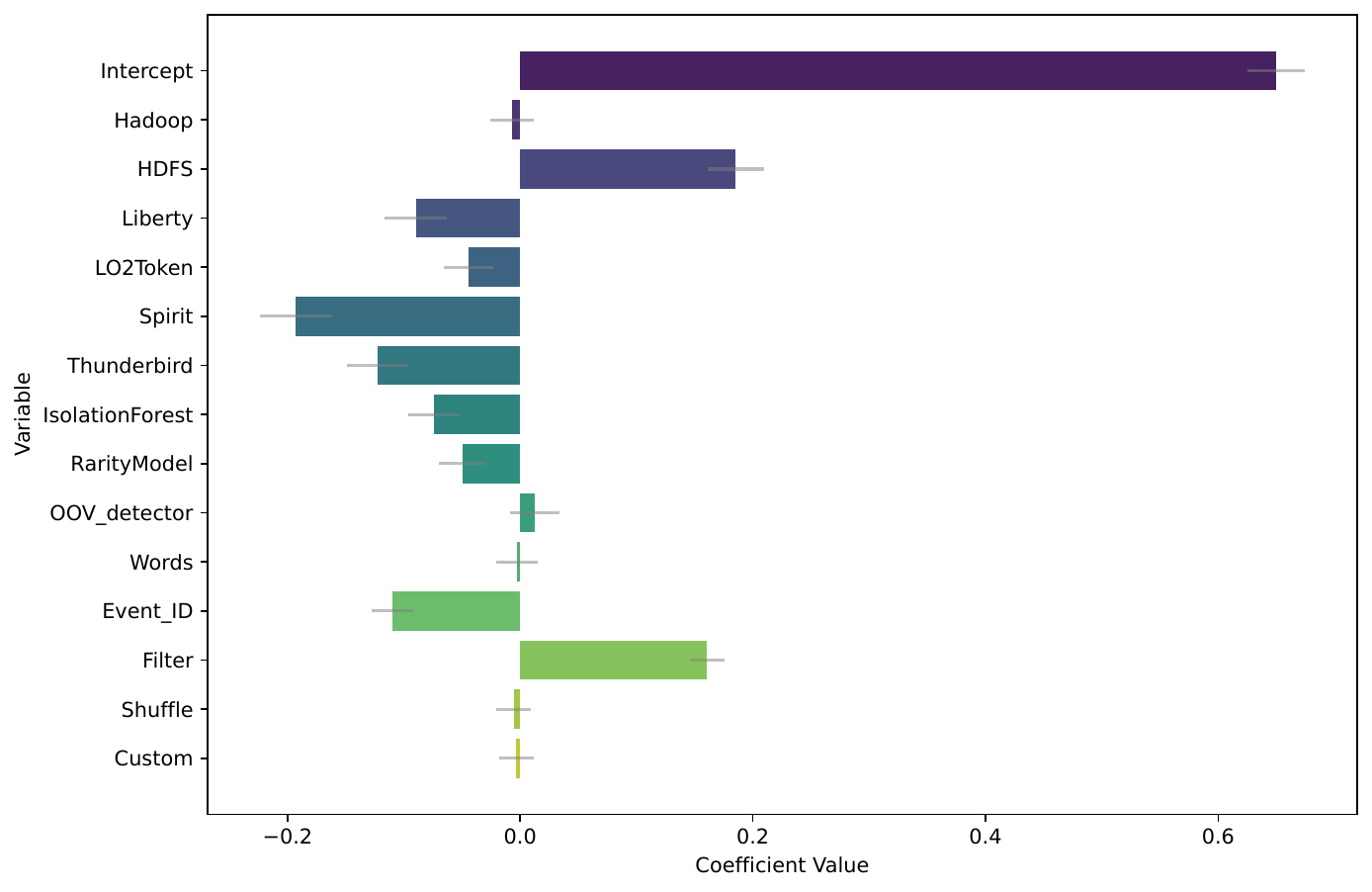}
    \caption{GLM with independent coefficients}
    \label{fig:coef_simple}
\end{figure*}

Figure \ref{fig:coef_simple} shows the coefficients with GLM using the variables independently. The Cox and Snell pseudo-R-squared for the model is 0.37 which means that the model  explains 37 \% of the variability. The purpose of GLM in our study is to give a high level overview of each variable when all the other variables are controlled. The independent variables suggest the following: HDFS is the easiest dataset, followed by BGL and Hadoop that are very close to each other. Sprit is the most difficult dataset. KMeans and OOVD are the best performing models while Isolation Forrest is the worst. Trigrams and Word representations beat the parsed Event IDs. Filtering anomalies from training data yields great benefits, whereas shuffling and using a custom loader does not have a significant impact. These findings are in line with the previous boxplots. 

\begin{figure*}
    \centering
    \includegraphics[width=0.98\linewidth]{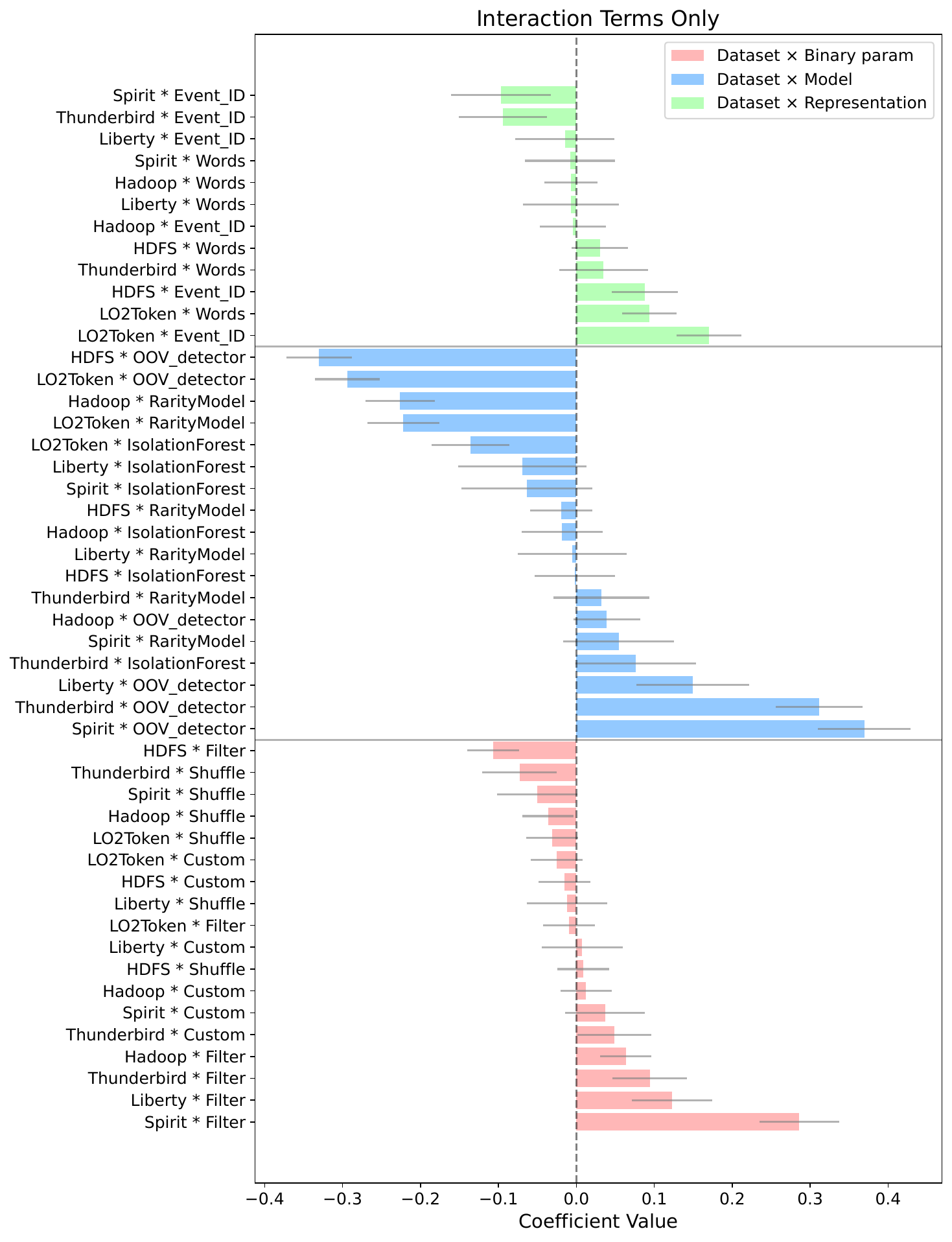}
    \caption{Coefficients of GLM with interaction terms}
    \label{fig:coef_interaction}
\end{figure*}

\begin{table*}
\centering
\caption{GLM Summary (with Interaction Terms)}
\label{tab:coef_estimates}
\small
\begin{tabular}{lrrrr}
\toprule
\textbf{Variable} & \textbf{Coef} & \textbf{Std Err} & \textbf{z} & \textbf{p-value} \\
\midrule
\textit{Interaction: Dataset $\times$ Model} & & & & \\
Hadoop $\times$ Isolation Forest & -0.0182 & 0.026 & -0.693 & 0.489 \\
HDFS $\times$ Isolation Forest & -0.0018 & 0.026 & -0.070 & 0.944 \\
Liberty $\times$ Isolation Forest & -0.0692 & 0.042 & -1.650 & 0.099 \\
LO2Token $\times$ Isolation Forest & -0.1361 & 0.025 & -5.364 & $< 0.001$ \\
Spirit $\times$ Isolation Forest & -0.0635 & 0.043 & -1.487 & 0.137 \\
Thunderbird $\times$ Isolation Forest & 0.0762 & 0.039 & 1.934 & 0.053 \\
Hadoop $\times$ Rarity Model & -0.2262 & 0.023 & -10.019 & $< 0.001$ \\
HDFS $\times$ Rarity Model & -0.0192 & 0.020 & -0.946 & 0.344 \\
Liberty $\times$ Rarity Model & -0.0054 & 0.036 & -0.152 & 0.879 \\
LO2Token $\times$ Rarity Model & -0.2220 & 0.024 & -9.380 & $< 0.001$ \\
Spirit $\times$ Rarity Model & 0.0544 & 0.036 & 1.497 & 0.134 \\
Thunderbird $\times$ Rarity Model & 0.0319 & 0.031 & 1.016 & 0.309 \\
Hadoop $\times$ OOV Detector & 0.0390 & 0.022 & 1.788 & 0.074 \\
HDFS $\times$ OOV Detector & -0.3302 & 0.021 & -15.404 & $< 0.001$ \\
Liberty $\times$ OOV Detector & 0.1494 & 0.037 & 4.047 & $< 0.001$ \\
LO2Token $\times$ OOV Detector & -0.2938 & 0.021 & -13.913 & $< 0.001$ \\
Spirit $\times$ OOV Detector & 0.3694 & 0.030 & 12.167 & $< 0.001$ \\
Thunderbird $\times$ OOV Detector & 0.3116 & 0.028 & 10.943 & $< 0.001$ \\
\midrule
\textit{Interaction: Dataset $\times$ Representation} & & & & \\
Hadoop $\times$ Words & -0.0070 & 0.017 & -0.402 & 0.687 \\
HDFS $\times$ Words & 0.0300 & 0.018 & 1.629 & 0.103 \\
Liberty $\times$ Words & -0.0069 & 0.032 & -0.219 & 0.826 \\
LO2Token $\times$ Words & 0.0939 & 0.018 & 5.279 & $< 0.001$ \\
Spirit $\times$ Words & -0.0082 & 0.030 & -0.276 & 0.782 \\
Thunderbird $\times$ Words & 0.0348 & 0.029 & 1.198 & 0.231 \\
Hadoop $\times$ Event ID & -0.0044 & 0.022 & -0.201 & 0.841 \\
HDFS $\times$ Event ID & 0.0878 & 0.022 & 4.050 & $< 0.001$ \\
Liberty $\times$ Event ID & -0.0146 & 0.032 & -0.450 & 0.653 \\
LO2Token $\times$ Event ID & 0.1701 & 0.021 & 7.967 & $< 0.001$ \\
Spirit $\times$ Event ID & -0.0969 & 0.033 & -2.966 & 0.003 \\
Thunderbird $\times$ Event ID & -0.0947 & 0.029 & -3.283 & 0.001 \\
\midrule
\textit{Interaction: Dataset $\times$ Preprocessing option} & & & & \\
Hadoop $\times$ Anomaly Free & 0.0635 & 0.017 & 3.780 & $< 0.001$ \\
HDFS $\times$ Anomaly Free & -0.1065 & 0.017 & -6.280 & $< 0.001$ \\
Liberty $\times$ Anomaly Free & 0.1227 & 0.026 & 4.654 & $< 0.001$ \\
LO2Token $\times$ Anomaly Free & -0.0093 & 0.017 & -0.545 & 0.585 \\
Spirit $\times$ Anomaly Free & 0.2861 & 0.026 & 10.942 & $< 0.001$ \\
Thunderbird $\times$ Anomaly Free & 0.0942 & 0.024 & 3.881 & $< 0.001$ \\
Hadoop $\times$ Shuffled & -0.0365 & 0.017 & -2.170 & 0.030 \\
HDFS $\times$ Shuffled & 0.0089 & 0.017 & 0.524 & 0.600 \\
Liberty $\times$ Shuffled & -0.0121 & 0.026 & -0.457 & 0.648 \\
LO2Token $\times$ Shuffled & -0.0313 & 0.017 & -1.842 & 0.065 \\
Spirit $\times$ Shuffled & -0.0502 & 0.026 & -1.921 & 0.055 \\
Thunderbird $\times$ Shuffled & -0.0730 & 0.024 & -3.006 & 0.003 \\
Hadoop $\times$ Custom & 0.0123 & 0.017 & 0.731 & 0.465 \\
HDFS $\times$ Custom & -0.0152 & 0.017 & -0.897 & 0.370 \\
Liberty $\times$ Custom & 0.0076 & 0.026 & 0.287 & 0.774 \\
LO2Token $\times$ Custom & -0.0253 & 0.017 & -1.491 & 0.136 \\
Spirit $\times$ Custom & 0.0370 & 0.026 & 1.415 & 0.157 \\
Thunderbird $\times$ Custom & 0.0485 & 0.024 & 2.000 & 0.046 \\
\bottomrule
\end{tabular}
\end{table*}

The model with interaction terms gives a much more nuanced view of the data. Figure \ref{fig:coef_interaction} shows the GLM coefficients when each other variable uses dataset as the interaction term. It also shows the confidence interval as a line on top of the bar to illustrate whether the association is statistically significant. The model has the pseudo-R-squared value of 0.70 which shows notable improvement from the previous model. 
For more detailed statistics on the GLM with interaction terms, see Table \ref{tab:coef_estimates}. It includes the coefficients, standard error, z-statistic, and the p-value.

The most interesting findings relate to the coefficients that clearly contradict the findings related to the independent coefficients, for instance, the independent result suggesting that parsing (EventID in the figure) degrades performance. With the interaction model, this effect appears to be dataset-specific. In the first block of Figure \ref{fig:coef_interaction} , the use of parsed EventIDs in the Spirit, Thunderbird, and Thunderbird-in datasets results in a decline in performance on average. Conversely, in the LO2 token and HDFS datasets, utilizing parsing yields a performance improvement. Thus, parsing exhibits the most significant impact, both positively and negatively.

Looking at the anomaly detection model section of the figure, represented in blue, a similar pattern emerges with the OOVD detector. OOVD performs exceptionally well in the Liberty, Thunderbird, and Spirit datasets, delivering substantial improvements. The high coefficients of OOVD for the supercomputer datasets highlight how dependent the model is on finding new, different log lines in the test data for those datasets.  However, it significantly degrades performance in the LO2 token and HDFS datasets. Thus, OOVD proves to be both the most effective and the least effective approach, depending on the dataset.

Finally, examining the red section of the figure, data filtering to be anomaly free appears to enhance performance across several datasets specifically Hadoop, Thunderbird, Liberty, and Spirit—while it negatively impacts the HDFS dataset.



Regarding the representation, the dataset-specific differences remain modest between Trigrams and Words. However, unlike with the independent variables, Event ID has a higher coefficient as an interaction term with HDFS and LO2Token datasets. It intuitively makes sense that the event ID would work better with sequence-based datasets, as its very purpose is to condense the log line into a single identifier. These results do not yet conclusively indicate that Event IDs should always be used with sequence-based datasets. 

Notably, the supercomputer datasets (Spirit, Thunderbird, and Liberty) exhibit consistent behavior, improving or degrading performance collectively. This suggests that these supercomputer datasets share similar properties which was to be expected due to their common source.

\subsection{GLM predictions analyzed}

\begin{figure*}
    \centering
    \includegraphics[width=\linewidth]{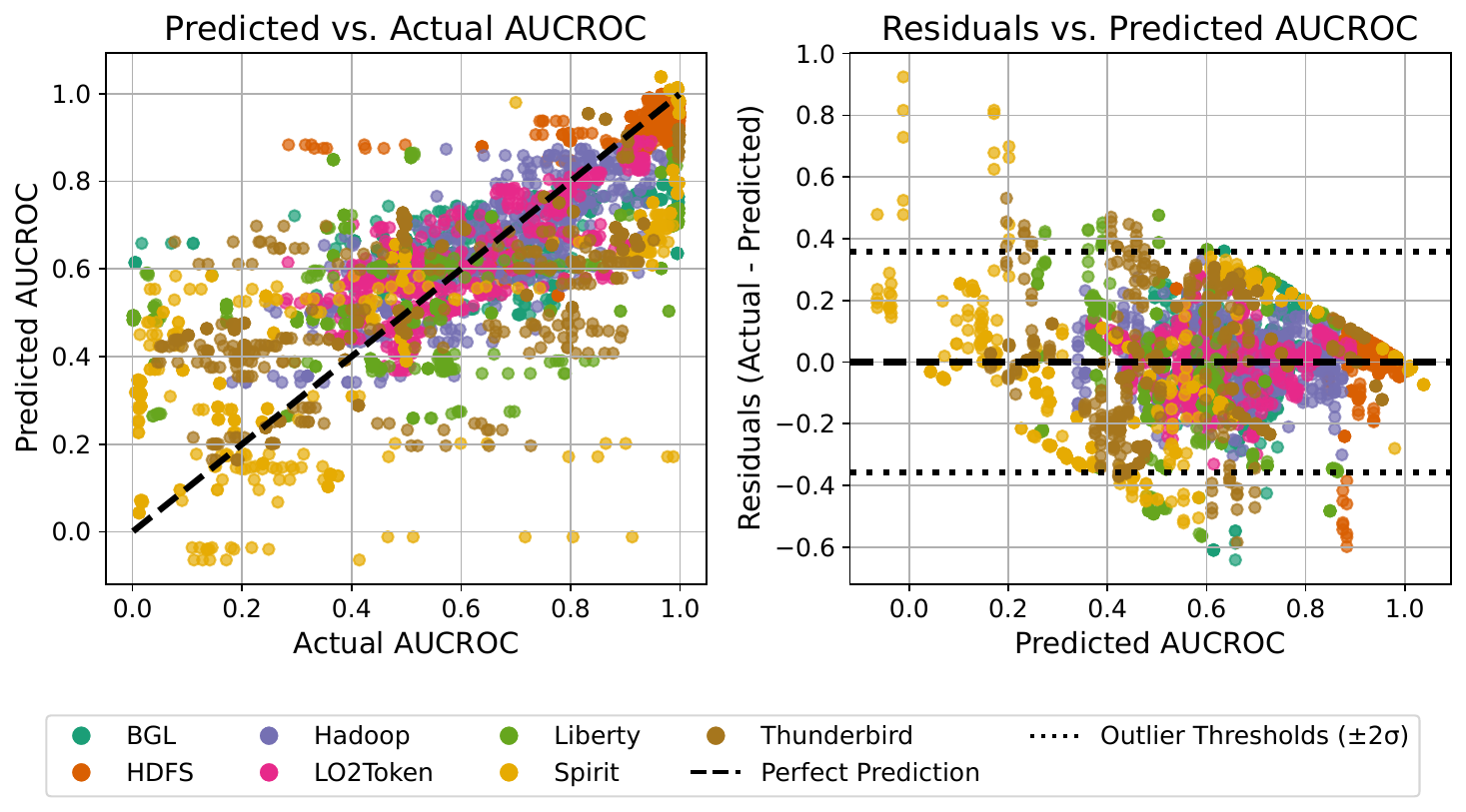}
    \caption{GLM predicted vs. actual, and residuals}
    \label{fig:residuals}
\end{figure*}

Figure \ref{fig:residuals} plots the actual AUCROC values in the data with the values predicted by the GLM model, effectively illustrating the residuals of the predictions. When the residuals are plotted against the predicted AUCROC, we can determine the upper and lower thresholds for the outliers, which are defined as being two standard deviations from the mean. Outliers are further classified in Table \ref{tab:outliers} to show instances where the prediction of the model fails. Examining these outliers is important because it helps identify the variability in the data that the regression model itself cannot capture. In particular, in the \textit{Residuals vs. Predicted AUCROC} plot in Figure \ref{fig:residuals} we can observe a diagonal trend suggesting a systematic problem in the model. Because it trends downwards, it means that the low predictions were actually higher and the high predictions were actually lower. Although we could adjust the model by increasing interaction terms, for this study we did not do that because we want to keep the results interpretable.

For practitioners, the residual analysis also suggests that the conclusions derived from the regression model are not necessarily generalizable to instances with a high number of outliers. For instance, the setup \textit{f1\_s0\_c1} has 0 positive outliers and 31 negative ones, which means that in 31 cases the model predicted a much better result than was actually the case. The prior analysis on the boxplots (Figure \ref{fig:boxplots}) suggests that these outliers originate from Liberty, where this particular setup did not work well. Overall, the variable with the most outliers was Isolation Forest with 44 positive outliers and 65 negative ones. For example, the positive outliers are very apparent in the Thunderbird boxplots as the orange dots. 

\begin{table*}
\centering
\begin{tabular}{llrrr}
\hline
\textbf{Category} & \textbf{Value} & \textbf{\shortstack{Positive \\ Outliers \\ (n=60)}} & \textbf{\shortstack{Negative \\ Outliers \\ (n=100)}} & \textbf{\shortstack{Total \\ Outliers}} \\
\hline
Dataset          & Liberty        & 20                          & 37                          & 57                       \\
                 & Thunderbird    & 21                          & 19                          & 40                        \\
                 & Spirit         & 14                          & 24                          & 38                        \\
                 & BGL            & 5                          & 10                          & 15                        \\
                 & HDFS           & 0                          & 10                          & 10                        \\
\hline
Setup            & f0\_s0\_c1     & 31                          & 0                          & 31                        \\
                 & f1\_s1\_c1     & 5                          & 28                          & 33                        \\
                 & f1\_s0\_c1     & 0                          & 29                          & 29                        \\
                 & f0\_s1\_c0     & 13                          & 5                          & 18                        \\
                 & f1\_s0\_c0     & 0                          & 17                          & 17                        \\
                 & f1\_s1\_c0     & 0                          & 14                          & 14                        \\
                 & f0\_s1\_c1     & 5                          & 5                          & 10                        \\
                 & f0\_s0\_c0     & 6                          & 2                          & 8                        \\
\hline
Model            & IsolationForest & 44                         & 65                         & 109                       \\
                 & KMeans          & 6                         & 30                          & 36                        \\
                 & OOV\_detector   & 10                         & 5                          & 15                        \\
\hline
Representation   & Event\_ID & 22                     & 43                          & 65                       \\
                 & Words        & 15                         & 40                          & 55                        \\
                 & Trigrams     & 23                         & 17                          & 40                        \\
\hline
\end{tabular}
\caption{Outliers by category}
\label{tab:outliers}
\end{table*}

\subsection{Execution time}

To address the computational efficiency of different models and representations, we measured the execution time. The results are divided into three tables. Table \ref{tab:time_sequence} shows the model training and test times for each sequence-based dataset as paired with each representation and model. Table \ref{tab:time_event} shows the same information for event-based datasets. We use the term "Test time" rather than "Inference time", because we measure the time to do both, the anomaly scoring and the binary prediction. This, combined with the fact that our training data is only 5\% of the total data (except for Hadoop which is 50\%), results in test times being generally much larger than training times. 

The results show that OOVD, KMeans and RM are very fast in each scenario. The difference between these models is often under a second, and even with the largest datasets, under 3 seconds. With those models, the slowest combination is using KMeans on Thunderbird with trigram representation, but even that is only about 10 seconds in total. With IF, the slowest model in this study, the same combination takes over 3.5 minutes.

Table \ref{tab:pre_time} contains the preprocessing times per dataset. Each setup goes through the normalization step, but only one of the representations is selected. After that, there is the "Split and Vectorize" step which has an execution time heavily impacted by the chosen representation. In the table, this is indicated by a letter: words (W), trigrams (T), and parsed event IDs (P). Overall, preprocessing is computationally demanding compared to the efficient anomaly detection models selected for this study. For example, if you were to select the fastest way to analyze BGL, the preprocessing would take around 47 seconds, while the model training and test would be under 4 seconds. Similar pattern applies to other datasets as well.  

The time to split and vectorize is dependent on the number of items in the data. Therefore, if we condense the logline into a single ID, the time is much lower than for the word representation. Conversely, if we split the line into much smaller trigrams, the time increases. However, parsing the data is much slower than splitting for words or trigrams. When combining both the operation to generate the representation, and to split and vectorize it, the results are clear. Word representation is by far the fastest with the average time of 30.92 seconds. Parsing is second with the time 104.33 seconds. Trigrams is the slowest with the time 185.89 seconds.  

\begin{table*}
\centering
\caption{Training and testing time for sequence datasets (seconds)}
\label{tab:time_sequence}
\begin{tabular}{ll|rr|rr|rr}

\hline
\textbf{Log rep.} & \textbf{Model} & \multicolumn{2}{c|}{\textbf{Hadoop}} & \multicolumn{2}{c|}{\textbf{HDFS}} & \multicolumn{2}{c}{\textbf{LO2Token}} \\
 & & Train & Test & Train & Test & Train & Test \\
\hline
\textbf{Words} & OOVD & $<$0.01 & $<$0.01 & $<$0.01 & 0.41 & $<$0.01 & $<$0.01 \\
 & IF & 0.06 & 0.04 & 0.38 & 6.55 & 0.07 & 0.08 \\
 & KMeans & 0.08 & $<$0.01 & 0.11 & 0.55 & 0.09 & 0.01 \\
 & RM & $<$0.01 & $<$0.01 & $<$0.01 & 0.43 & $<$0.01 & $<$0.01 \\
\hline
\textbf{Trigrams} & OOVD & $<$0.01 & $<$0.01 & $<$0.01 & 0.49 & $<$0.01 & $<$0.01 \\
 & IF & 0.07 & 0.02 & 2.74 & 26.10 & 0.10 & 3.11 \\
 & KMeans & $<$0.01 & $<$0.01 & 0.14 & 1.19 & 0.02 & 0.06 \\
 & RM & $<$0.01 & $<$0.01 & 0.01 & 0.50 & $<$0.01 & 0.01 \\
\hline
\textbf{Events} & OOVD & $<$0.01 & $<$0.01 & $<$0.01 & 0.40 & $<$0.01 & $<$0.01 \\
 & IF & 0.08 & $<$0.01 & 0.53 & 3.67 & 0.24 & 0.03 \\
 & KMeans & $<$0.01 & $<$0.01 & 0.01 & 0.48 & $<$0.01 & $<$0.01 \\
 & RM & $<$0.01 & $<$0.01 & $<$0.01 & 0.40 & $<$0.01 & $<$0.01 \\
\hline

\end{tabular}
\end{table*}

\begin{table*}
\centering
\caption{Training and testing time for event datasets (seconds)}
\label{tab:time_event}
\begin{tabular}{ll|rr|rr|rr|rr}

\hline
\textbf{Log rep.} & \textbf{Model} & \multicolumn{2}{c|}{\textbf{BGL}} & \multicolumn{2}{c|}{\textbf{Tb}} & \multicolumn{2}{c|}{\textbf{Liberty}} & \multicolumn{2}{c}{\textbf{Spirit}} \\
 & & Train & Test & Train & Test & Train & Test & Train & Test \\
\hline
\textbf{Words} & OOVD & $<$0.01 & 3.63 & $<$0.01 & 5.55 & $<$0.01 & 4.46 & $<$0.01 & 4.41 \\
 & IF & 0.64 & 24.98 & 1.08 & 49.07 & 0.92 & 30.85 & 1.19 & 37.80 \\
 & KMeans & 0.16 & 3.96 & 0.26 & 6.48 & 0.19 & 4.86 & 0.22 & 5.26 \\
 & RM & $<$0.01 & 3.63 & 0.01 & 5.89 & $<$0.01 & 4.40 & $<$0.01 & 4.43 \\
\hline
\textbf{Trigrams} & OOVD & $<$0.01 & 3.75 & $<$0.01 & 6.96 & $<$0.01 & 4.54 & $<$0.01 & 4.82 \\
 & IF & 2.29 & 59.41 & 4.67 & 212.76 & 3.44 & 71.76 & 4.63 & 96.91 \\
 & KMeans & 0.23 & 4.91 & 0.84 & 9.40 & 0.29 & 5.97 & 0.41 & 7.17 \\
 & RM & 0.02 & 3.74 & 0.05 & 6.53 & 0.02 & 4.63 & 0.03 & 5.14 \\
\hline
\textbf{Events} & OOVD & $<$0.01 & 3.50 & $<$0.01 & 5.96 & $<$0.01 & 4.30 & $<$0.01 & 4.55 \\
 & IF & 0.46 & 22.64 & 0.93 & 30.80 & 0.55 & 25.17 & 0.66 & 27.03 \\
 & KMeans & 0.05 & 3.75 & 0.07 & 6.54 & 0.05 & 4.58 & 0.06 & 4.89 \\
 & RM & $<$0.01 & 3.52 & $<$0.01 & 6.29 & $<$0.01 & 4.30 & $<$0.01 & 4.56 \\
\hline

\end{tabular}
\end{table*}

\begin{table*}
\centering
\caption{Pre-processing times (seconds)}
\label{tab:pre_time}
\begin{tabular}{l|rrrrrrr|r}
\hline
\textbf{Time to...} & \textbf{Hadoop} & \textbf{BGL} & \textbf{HDFS} & \textbf{Tb} & \textbf{Liberty} & \textbf{Spirit} & \textbf{LO2T.} & \textbf{Avg} \\
\hline
Normalize & 1.16 & 16.58 & 25.20 & 27.94 & 15.36 & 21.37 & 13.01 & 17.23 \\
Create trigrams & 0.63 & 10.12 & 13.16 & 21.67 & 10.62 & 13.91 & 9.61 & 11.39 \\
Create words & 0.04 & 0.71 & 0.80 & 1.70 & 0.82 & 1.20 & 0.62 & 0.84 \\
Parse event IDs & 1.67 & 34.58 & 70.43 & 265.89 & 89.34 & 155.65 & 25.24 & 91.83 \\
Split \& Vec. (W) & 0.45 & 29.44 & 30.85 & 59.80 & 34.75 & 45.35 & 9.91 & 30.08 \\
Split \& Vec. (T) & 4.98 & 149.64 & 251.89 & 353.65 & 158.24 & 208.70 & 94.43 & 174.50 \\
Split \& Vec. (P) & 0.16 & 14.73 & 7.36 & 27.95 & 16.71 & 18.06 & 2.52 & 12.50 \\
\hline
\end{tabular}
\end{table*}

\subsection{Supplementary analysis: Shuffling on a supervised model}
\label{sec:supplementary}

\begin{figure*}
    \centering
    \includegraphics[width=\linewidth]{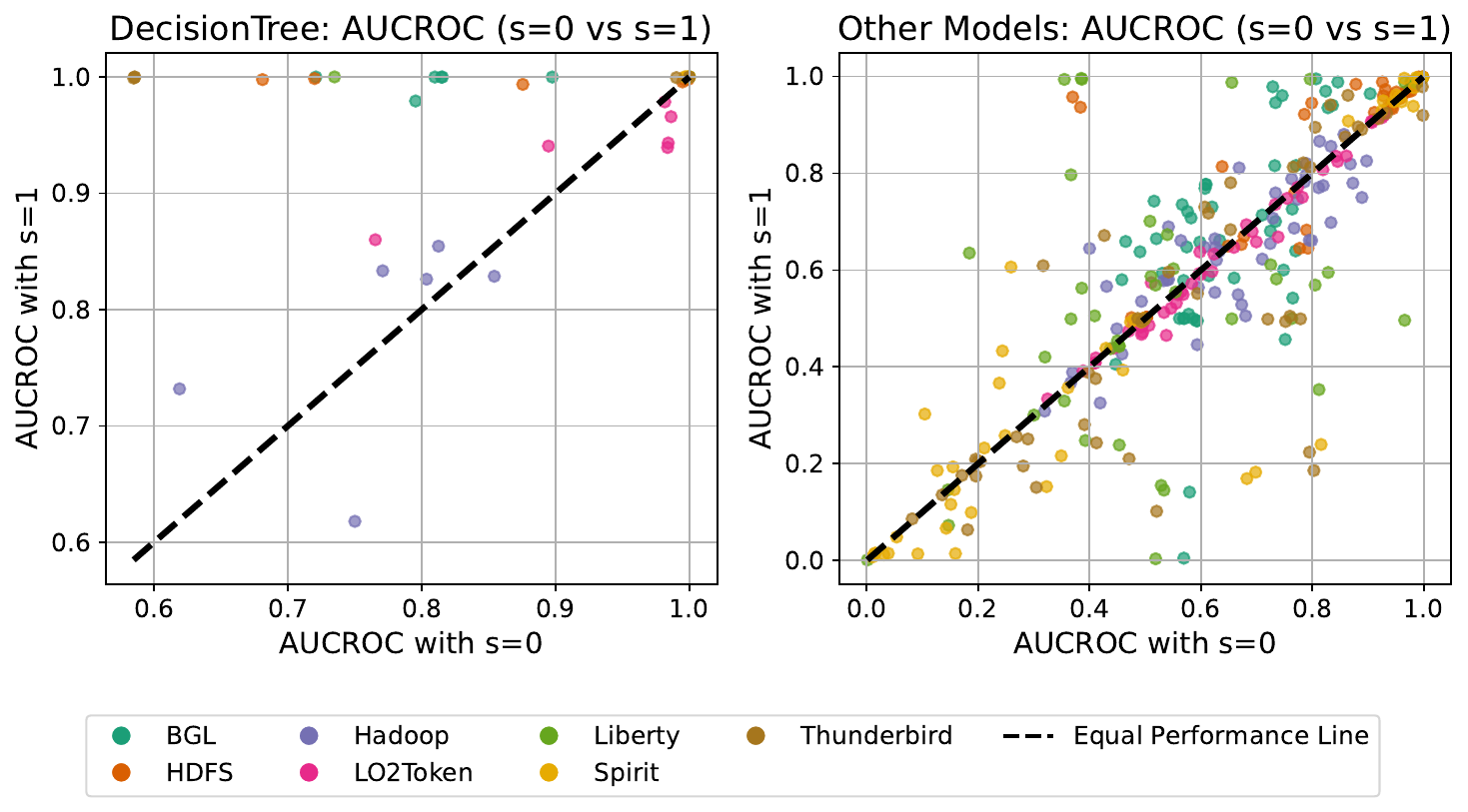}
    \caption{Metrics plotted with s = 0 against s = 1}
    \label{fig:suffleplots}
\end{figure*}

The most surprising finding was the small impact of the \textit{Shuffle} preprocessing option. Intuitively, shuffling the data before assigning training and test sets should always have a positive effect, as it provides more diverse training data and allows a loop hole to see to the future. To address this issue further, we ran the experimental setups with a supervised Decision Tree (DT) classifier. The assumption is that supervised models are inherently more dependent on diverse training data due to the use of labels. The plot \ref{fig:suffleplots} shows the scores when plotting without shuffling (s~=~0) against with shuffling (s~=~1) for runs with otherwise the same parameters. This effectively shows that dots above the black line perform better with shuffling. For DT we can observe that most datapoints overlap at perfect accuracy but for the ones that do not, they are generally above the line. The same does not apply for other models. There are some instances where shuffling clearly improves performance, such as the green dots representing Liberty. These correspond to the same findings we have presented in the boxplot and in the residual analysis. 

To get a better idea of the general impact of shuffling we calculated the averages and standard deviations between s~=~0 and s~=~1 for DT, other models, and other models when comparing only runs with anomalous training data (f~=~0). These are presented in table \ref{tab:shuffling_impact}. Because DT requires samples from both classes (normal and anomaly), we included the f~=~0 row for other models as well. 

The results are clear; supervised models really benefit from the more diverse training data while for unsupervised models it is mostly negligible. For unsupervised models, these results suggest that even with a small amount of training data from the beginning, the models can roughly learn what normal behavior looks like without needing exact samples.

\begin{table*}
    \centering
    \caption{Impact of Shuffling on Model Performance}
    \label{tab:shuffling_impact}
    \begin{tabular}{lccccc}
        \toprule
        \multirow{2}{*}{Model Type} & \multirow{2}{*}{Shuffle} & \multicolumn{2}{c}{AUCROC} & \multicolumn{2}{c}{F1 Score} \\
        \cmidrule(lr){3-4} \cmidrule(lr){5-6}
        & & Mean & Std & Mean & Std \\
        \midrule
        \multirow{2}{*}{Decision Tree} 
        & s=0 & 0.8667 & 0.1403 & 0.7893 & 0.2674 \\
        & s=1 & 0.9500 & 0.1071 & 0.9626 & 0.0680 \\
        \midrule
        \multirow{2}{*}{Other Models (All)} 
        & s=0 & 0.6246 & 0.2550 & 0.5985 & 0.2886 \\
        & s=1 & 0.6166 & 0.2816 & 0.6115 & 0.2932 \\
        \midrule
        \multirow{2}{*}{Other Models (f = 0 only)} 
        & s=0 & 0.5735 & 0.2465 & 0.5830 & 0.2920 \\
        & s=1 & 0.5523 & 0.2686 & 0.5731 & 0.3018 \\
        \bottomrule
    \end{tabular}
\end{table*}

\section{Discussion}

\textbf{RQ1: What effect do the model, representation and preprocessing options have on anomaly detection performance on software logs?}

Similarly to previous work \cite{speedandperf}, our results indicated that there is no single best option. For example, from the models, OOVD reached the best average AUCROC score, but it simply does not work unless you have clean training data. Even the question of whether to use the clean training data depends on the model and dataset. We suspect that this is due to the fact that the models cannot inherently handle out-of-vocabulary terms, which become much more prevalent if the anomalies are filtered from the training data. To be clear, there is a meaningful effect in the selection of the approach, but based on the results, it cannot be generalized. 

\textbf{RQ2: To what extent is the variability explained by the dataset?}

As expected, the effect of the dataset was so large that it is hard to generalize the findings regarding the other variables. As such, it is one of the main contributions of this study to present the characteristics of each dataset. For example, the event-level supercomputer logs are completely dependent on finding new log messages, as illustrated by the performance of the OOVD model on anomaly free data. For the sequence-based models this was not the case. Especially, the new LO2Token does not work with OOVD at all, because it has a small number of anomalous log messages in normal sequences. We expect the same to be true for HDFS and Hadoop, to some extent.

\textbf{RQ3: How do the computational costs of different models, representations, and preprocessing approaches compare?}

For practical purposes, the results show that word representation with OOVD, KMeans or RM is the fastest option. To improve the performance further, emphasis should be placed on improving the preprocessing practices rather than the models. The results are mostly in-line with previous work \cite{speedandperf}. The absolute values differ between these studies because of a different execution environment, but the relative differences remain similar. Compared to the previous study, there was a major improvement on Trigram generation time. This was due to a new function we implemented that generates the trigrams by splitting on three positions rather than iteratively. It still remains the slowest option due to the amount of items to be vectorized. While comparing execution times between studies is difficult, overall the results for execution time presented in this paper remain much lower than in previous papers for both preprocessing \cite{XDrain} and anomaly detection \cite{expreport}.

\begin{figure*}
    \centering
    \includegraphics[width=\linewidth,height=0.95\textheight,keepaspectratio]{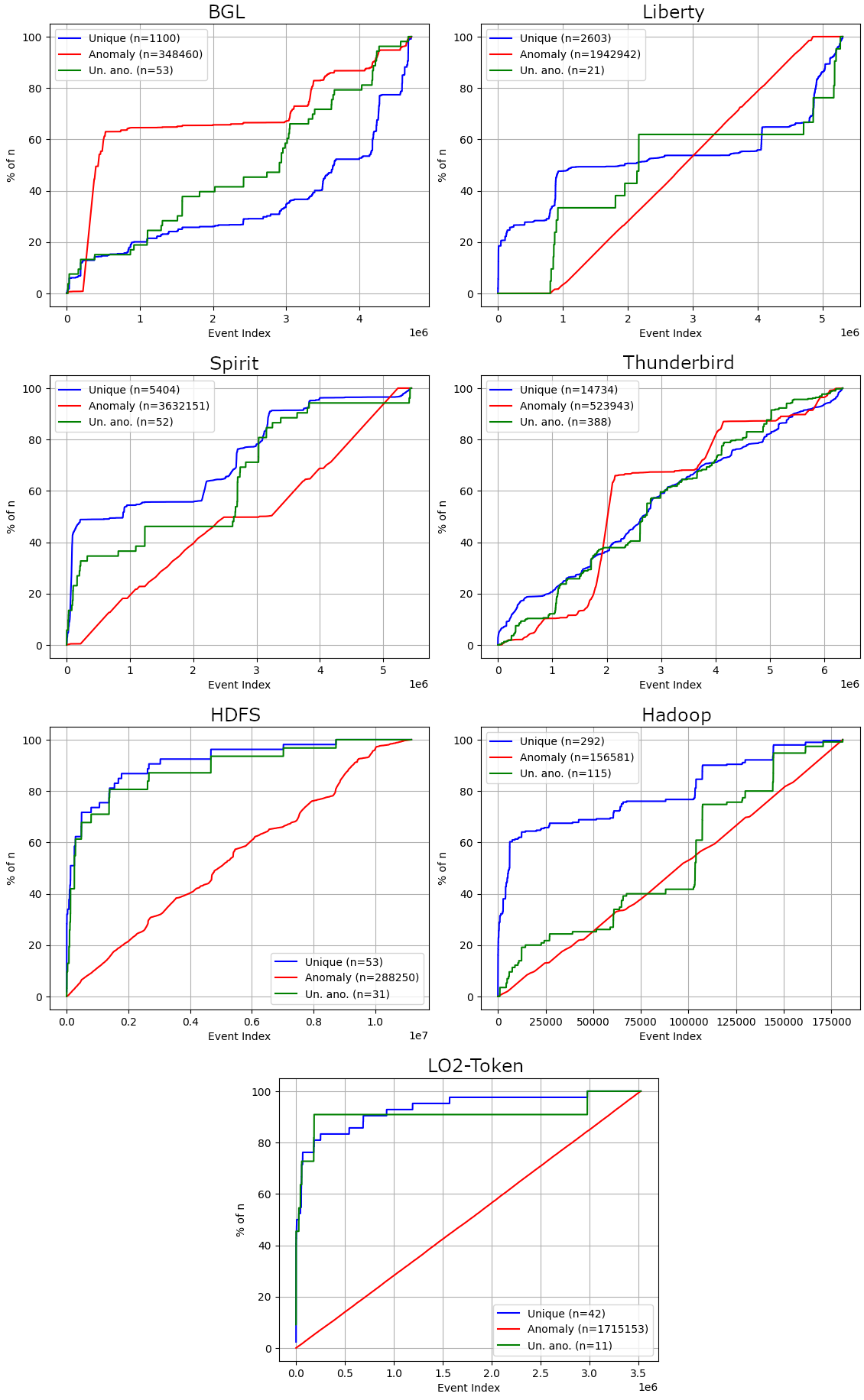}
    \caption{Cumulative sum of events by dataset}
    \label{fig:cumsum}
\end{figure*}

 The use of random instead of chronological random training data selection should have improved the anomaly detection performance \cite{howfar}, yet our experiments showed negative or no effect. We suspect that the difference between our results and the previous results is that we used a much smaller portion of the data for training. Assuming you reach good coverage of the different log messages, then it is better to find them throughout the data, but if the number of messages to be included is low, then it might yield better results to use the chronologically early ones. Figure \ref{fig:cumsum} illustrates the cumulative sum of unique events (parsed with Drain), anomalous events, and unique anomalous events in the datasets. As we can see, many datasets have a surge of new unique events in the beginning. If we use chronological selection, all events after a given point will go to the test set and are therefore “new” as the model encounters them. To perform well, the model needs to find similarities between the new normal events and the known ones and assign them correctly as normal. The results suggest that the unsupervised models are able to do this reasonably well. Such figures could be used by practitioners to highlight relevant parts of execution where there is sudden surge of new activity, and then make informed decisions on whether or not they should be included in the training data. The plots also show that the sequence-based datasets generally reach a high percentage of all unique events relatively quickly. This indicates that the sequences themselves have a lot of common log messages which should make the anomaly detection process easier.

The focus of this study was on the examination of different combinations of experimental variables rather than proposing a new approach. Therefore, we do not want to make direct comparisons with the approaches used in other papers. It would be very challenging if not impossible to create fair comparisons, as most papers do not report all the necessary details regarding preprocessing that were crucial for our study. In general, the results presented in this study align with the findings of other studies. 

This study opens various directions for future research. For instance, one could study qualitatively what kind of log messages cause a certain model not to perform well. We did it for LO2Token and OOVD as mentioned in Section~\ref{sec:regressionanalysis} because we knew the dataset well. However, it was not within the scope of the paper to include similar analysis for all datasets. In the previous study \cite{speedandperf}, we recommended developing an algorithm to combine OOVD with other models as they do not inherently handle OOV terms well. This requires a technical solution that was not in the scope of this paper.   

While one of the goals of this paper was to improve our understanding of how different factors affect the anomaly detection performance, it is important to note that our research method does not address this explicitly. To be precise, the research question asked \emph{what} is the effect of the factors rather than \emph{how} do the factors affect each other. Empirical experiments are rooted in positivist or post-positivist traditions that aim to establish that there exist a causal connection between two factors but it cannot address why or how it exists. As such our interpretations of \emph{why} certain settings work well in a given scenario are only interpretations.    

\subsection{Threats to validity}

This section describes the threats to validity as divided into internal validity, external validity, and construct validity. 

\subsubsection{Internal validity}
Internal validity considers to what extent does study design and methodology allow us to draw accurate conclusions about the factors being studied. A key limitation is that we did not optimize the model-specific parameters. This means that the performance metrics that we report may not reflect the best possible performance for each approach. Furthermore, it likely influences the possibility to conduct fair comparative analysis between methods. We chose not to include hyperparameter optimization because the number of factors being studied was already so large and hyperparameter optimization would increase that number exponentially. 

\subsubsection{External validity}
External validity addresses the generalizability of our findings in other contexts and practical applications.  One limitation is the applicability of our custom loader implementations. We chose to use the loaders to extract only the log message for two reasons. First, this approach provides a uniform and fair baseline for comparing the datasets, since the fields in the log line may differ between datasets, but all include the log message. Second, log parsing as a discipline focuses predominantly on working with just the log messages. However, we acknowledge that using only the log message likely does not yield the best results for all datasets. In practical applications, practitioners might implement loaders tailored to each case to retain other relevant data while discarding irrelevant fields. For our setup, this was not feasible, as the fairness of the evaluation would have been jeopardized. Nevertheless, we suspect that the results regarding the use of custom loaders would have been better under such configurations.

\subsubsection{Construct validity}
Construct validity concerns whether we are measuring what we are supposed to. A limitation in our study is that using ground-truth labels to compute the F1-score is not realistic for a practical setting, because such labels for anomaly detection are typically unavailable in the industry. To overcome this we used AUC-ROC as our primary  The differences in evaluation strategy also means that the results are not directly aligned with those used in comparable prior work \cite{expreport}, \cite{loganomaly}.

\section{Conclusion}
First, we present one of the most comprehensive empirical evaluations of unsupervised anomaly detection approaches on software logs. We study the effects of four models (KMeans, IsolationForest, RarityModel, and Out-Of-Vocabulary Detector), three representations (words, trigrams, and parsed events), and three preprocessing options (anomaly-free training data, shuffled vs. chronological data selection, and custom vs. generic log message extraction), across seven datasets. This analysis, with 672 experimental setups, offers practitioners and researchers a detailed understanding of how these factors interact and influence performance.

Second, our findings highlight the critical role of dataset characteristics in determining the effectiveness of anomaly detection methods.  Our statistical analysis, using generalized linear models, quantifies the interactions between factors and datasets, revealing that the dataset itself explains the majority of performance variability. While no single approach universally outperforms others, we identify patterns that generalize across similar datasets. For example, log parsing improves performance on sequence-based datasets but can degrade it on event-based ones. Similarly, the Out-Of-Vocabulary Detector excels on supercomputer datasets with anomaly-free training data but fails on sequence-based datasets like HDFS and LO2Token. These insights underscore the importance of tailoring the choice of model, representation, and preprocessing to the specific properties of the dataset at hand.

Third, we challenge the assumption that shuffling training data—a form of data leakage—leads to unrealistic performance improvements. Our results show that while shuffling benefits supervised models, its impact on unsupervised models is negligible. 

Fourth, we provide evidence that using anomaly-free training data improves  performance for unsupervised models. While getting anomaly-free data requires additional effort, in practice, this is not as difficult as one might assume. For example, anomaly-free data can be obtained from periods of normal operation—such as days when site reliability engineers confirm that system performance was stable—which can then be used for model training and deployed in production.

For new datasets, we recommend first identifying similar datasets in the literature and adopting the methods that have proven effective in those contexts. When no comparable datasets exist, empirical experimentation remains necessary.

\section*{CRediT authorship contribution statement}
 \textbf{Jesse Nyyssölä}: Writing – review \& editing, Writing – original draft, Methodology, Data curation, Conceptualization, Software, Formal analysis. 
 \textbf{Mika Mäntylä}: Writing – review \& editing, Writing – original draft, Methodology, Conceptualization, Supervision, Validation. 
 \textbf{Yuqing Wang}:  Writing – review \& editing, Validation.

\section*{Declaration of competing interest}
The authors declare that they have no known competing financial interests or personal relationships that could have appeared to influence the work reported in this paper.

\section*{Acknowledgment}
Funded by the Academy of Finland (grant ID 359861).

\section*{Data availability}
Scripts for replication, parsed pre-run results, and the LO2Token dataset are available on Zenodo: \url{https://doi.org/10.5281/zenodo.17787039}

\bibliographystyle{elsarticle-num}

\bibliography{ref}

\end{document}